\documentclass[twocolumn]{article}

\usepackage{graphicx}
\usepackage{dcolumn}
\usepackage{bm}
\usepackage{hyperref}

\begin{document}


\title{Acoustic emission signals resulting from the drying induced
  fractures of \emph{Phyllostachys Pubescens} bamboo, Evidence of
  scale free phenomena}
\author{ \small {Gabriel Villalobos   Universidad de Bogot\'a Jorge Tadeo Lozano. Carrera 4 N 22- 70. Bogot\'a, Colombia}\\
\small{  Computational Physics for Engineering Materials, IfB,ETH Z\"urich, Schafmattstr. 6, CH-8093 Zurich, Switzerland.  }}

\date{\today}
\maketitle

\begin{abstract}
I have performed experimental measurements of acoustic emission signals
resulting from the drying process of \emph{Phyllostachys Pubescens}
bamboo. The emphasis was on identifying individual events, and
characterize them according to their time span and energy release. My
results show a histogram of experimental squared voltage distributions
nicely fit into a power law with exponent of $-1.16$, reminiscent of
scale free phenomena. I have also calculated the average signal
shape, for different time spans of the system, and found an
asymmetrical form. The experimental evidence points to the system
having an isolated large crack at the beginning of the simulation. 
\end{abstract}


\section{Introduction}
Bamboo is a giant grass, with remarkable mechanical characteristics
\footnote{Values of the elastic modulus of the fiber cell wall of
  either 10.4 $GPa$ \cite{Zou20091375} or 16.1 $GPa$ \cite{YU07}have
  been reported in the literature}. Nonetheless, as it is a natural
material, its drying requires special care in order to prevent the
cracking due to dessication \cite{Montoya06}. In some aspects Bamboo
drying is similar to wood drying: there is a first fast drying stage,
in which free water is removed from the plant; followed by a slow
stage in which bond water, from within the cell walls, has to be
removed \cite{Yu08}. The main difference between drying this two
natural materials comes from the structure of their tissues. Unlike
wood, bamboo has a special direction, the culm, along which different
tissues are aligned: its fibers, conductive material and
parenchymatous tissues \cite{Janssen1981}.

Very few research groups have studied the relationship between
shrinkage and fracturing in bamboos.  Overall shrinkage of bamboo
canes can be attributed to the shrinkage of the fiber-phloem bundles,
as well as to the shrinkage of parenchyma tissue. In the latter, the
collapse of cells can be an important factor. Obataya et. al. have
shown how the slow drying process can lead to the collapse of
\textit{Arundo donax} cane, \cite{Obataya05A,Obataya05B}. Those authors
claim that an slow drying schedule increases the intensity of
collapse.

On a more general fracture case, Habibi and Lu have studied the path
followed by different cracks within the bamboo culm, by means of
micro indentation experiments as well as tensile tests
\cite{habibi2014crack}. Interestingly, the micro crack grows mainly in
the interface between the parenchyma cells. When the crack reaches a
fiber it is sometimes possible for it to continue within the fiber,
again in the interfaces between fiber cells. They have also shown how
the voids within the bamboo have an effect of crack grow deflection
and crack tip energy absorption.

I would expect a different path of growth for dessication induced fractures
compared to indentation; as drying stresses are more homogeneous, not
having a unique defined direction of deformation. My attempts to look
for the fracture in the surface proved futile; which may be caused by
fractures being smaller than the resolution of my optical microscope,
or by them being inside of the samples themselves
\cite{VILLA12}. Then, the problem calls for a different way to characterize
the fracture evolution during drying, so I turned the research focus into
acoustic emission. As fractures are created at every scale, some of
the energy is released in the form of an acoustic wave.

In the present letter I have characterized the dessication induced
fractures in bamboo \textit{Phyllostachys Pubescens}. To do so, I
performed drying experiments in which I recorded the acoustic signal
produced by the sample. Since it was not subject to any other strain, it is
safe to assume that the sounds are generated by the dessication
induced strains. The goal of this measurement is to obtain information
about the drying induced fractures. The hypothesis being that the
average shapes of fractures, the distributions of energy
released in an avalanche event, and also the waiting times between
avalanches; all have useful information about the process of the
drying induced fracturing.

The organization of the paper is as follows. In section 2 I describe
the experiments, materials and methods used. In Section 3 I will present
my results and conclusions.

\section{Materials and Methods}
Two kinds of experiments were performed, namely, on characterizing the
process of drying and on characterizing the acoustic emissions. For
the former I measured the weight as a function of drying time. For
the latter, I measured the acoustic emission resulting from the
drying process.
\begin{table}[htb]
  \begin{center}
    \begin{tabular}{ | l | p{2cm} | p{2cm} |}
      \hline
      Label & Number of Samples & Length \\ \hline
      Sample 0 & 4 & $1$ $cm$  \\
      Sample 1 & 1 & $10$ $cm$  \\
      Sample 2 & 10 & $10$ $cm$  \\
      \hline
    \end{tabular}
    \caption{\emph{Phyllostachys Pubescens} inter-node samples used in
      the present study.}
  \end{center}
\end{table}

\subsection{Materials}
Bamboo samples were prepared from a 4 year old culm
\emph{Phyllostachys Pubescens}, harvested on June 2011, from Friedrich
Eberts' plantation in Chiavari, Italy. Three sets of inter-node cut
samples were labeled and prepared, as shown in the table. On Sample $0$,
I performed drying experiments aimed to characterize the likelihood
of collapse and fracture. On Sample 1, I took the weight measurements
during drying, in order to characterize the speed of drying. Finally,
on Sample 2, I measured the acoustic emission of breaking events
during drying. Between measurements the samples were kept on a climate
controlled room, with relative humidity of 80\%, to minimize the
amount of drying happening before the oven treatment. I performed the
drying experiments using a SalvisLAB Pantatherm D oven.

\subsection{Methods}
Two sets of experiments were performed. First, for the
characterization of the speed of drying, $20$ inter-nodal bamboo
samples of $10$ $cm$ width were dried in the oven at a set temperature
of $100$ $C$. In this experiment the percentage of weight loss
of the  bamboo samples as function of time was monitored; and later averaged to build the plot.

The second set of experiments characterized the intensity of the
microscopic cracks by measuring their acoustic emission during
drying. The drying temperature was again $100$ $C$, and the
digitization was performed using a standard condenser microphone and
a digital sound card at a rate of 9600 bps. The voltage on the
microphone is proportional to the energy released as acoustic
waves. After the recording, the whole data set was screened to check
for noise coming not from the drying specimen.
\begin{figure}[htb]
\begin{center}
\resizebox{7cm}{!}{\includegraphics{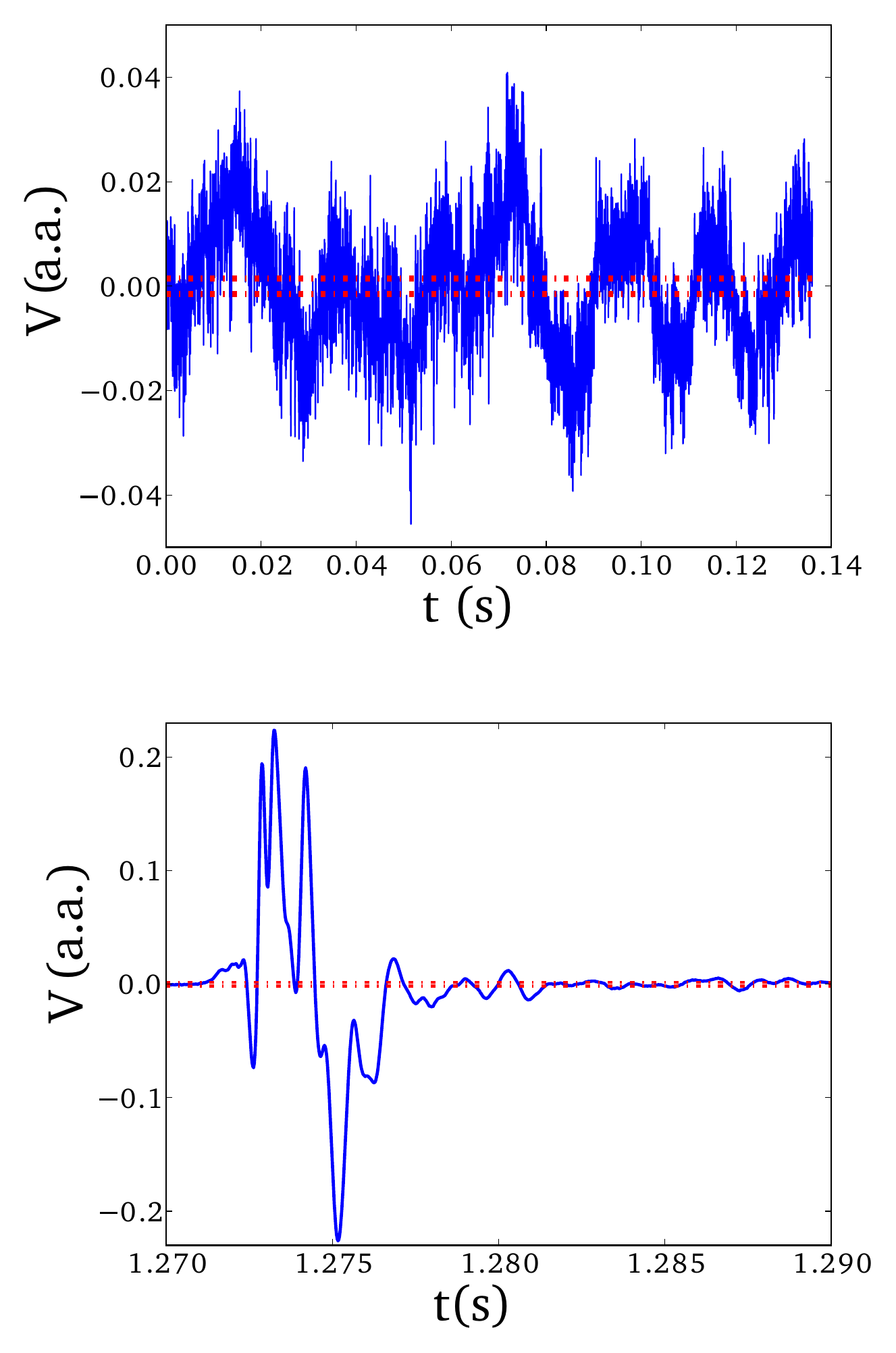}}
\caption{(Color Online). Examples of voltage on the microphone, blue
  continuous line (proportional to the sound intensity), and set noise
  threshold. (Upper) A  cascade of small sized breaking. (Lower) A single large
  crack. (Sound included as supplementary material.)}
\label{fig:voltagesample}
\end{center}
\end{figure}

\begin{figure}[htb]
\begin{center}
\resizebox{7cm}{!}{\includegraphics{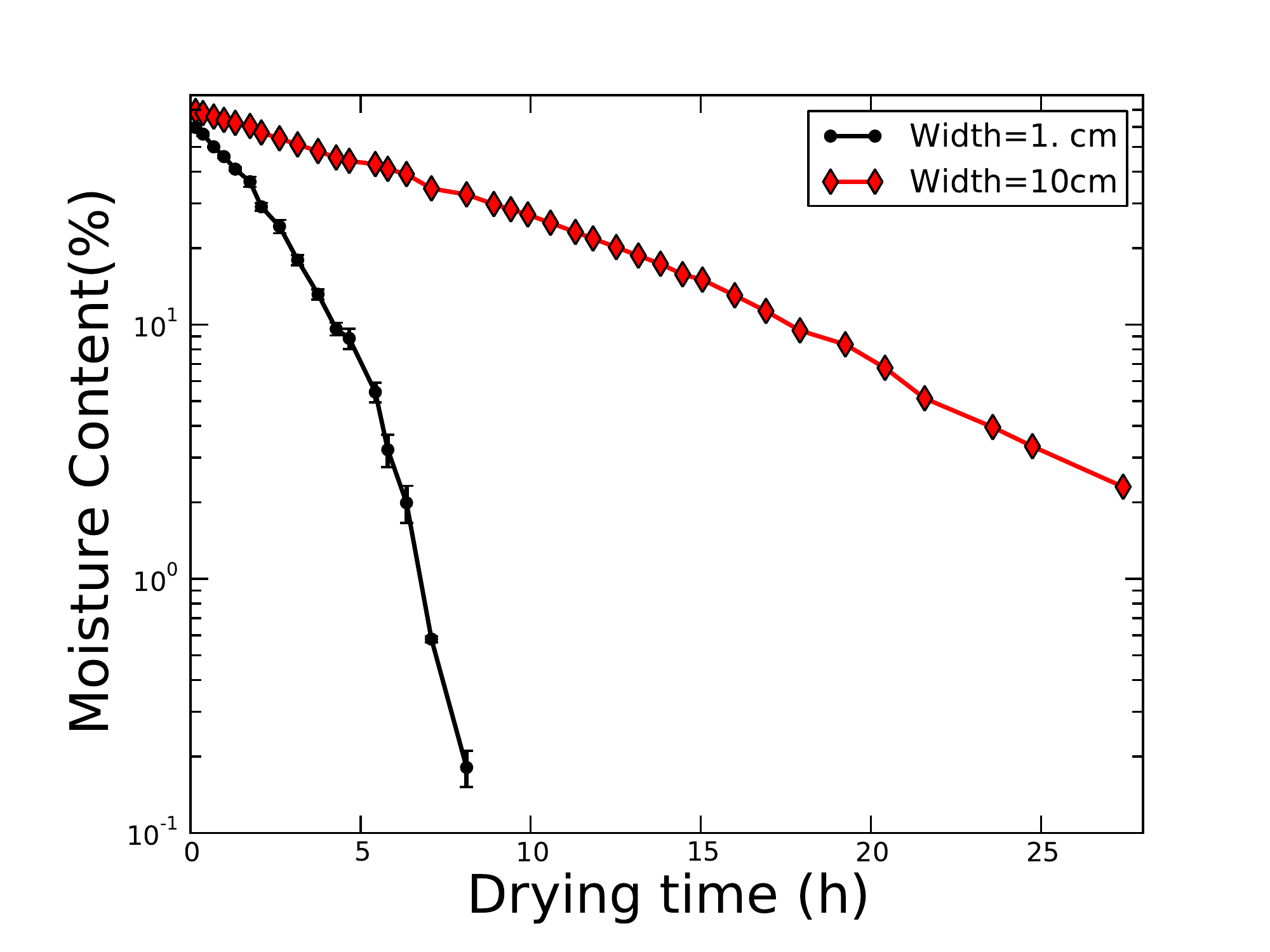}}
\caption{Moisture content as a function of temperature for: one 10 cm
  sample (diamonds) and four 1 cm samples (black point at the average
  value).}
\label{fig:dryingtimes}
\end{center}
\end{figure}
The recording took place during the first 3 hours of drying.
Background noise, both acoustic and from the electronics, was filtered
by means of setting a minimum threshold for recording a voltage signal
(See Fig. \ref{fig:voltagesample}). Experiments were performed
overnight, to reduce the amount of noise from the environment. Later
on, the recordings were split in $6$ $s$ samples; I listened to each
one of them in order to exclude the sections of the recording that
presented strong noises, probably arising from the oven. Finally, the
signals were converted in text files.

Now, I claim that  the continuous acoustic emission events can be interpreted as produced by
the breaking of bamboo structures; be them fibers, conductive
materials or parenchyma. Therefore they are be called indistinctly
crack events or acoustic events. A single event is defined as a
continuous signal that lies above the minimum recording threshold.

Since my objective is the characterization of the fracture process,
the exact relationship between the voltage
measured by the microphone and the amplitude of the original sound
wave in $dB$ does not need to be explicitly known. Moreover, this calibration is a quite complex process,
that involves the electronics of the sound card and the software of
the computer used to the recording (OS X, a proprietary and closed
source operating system by Apple Inc.). Therefore I will report my
experiments in arbitrary units of amplitude: $a.a.$, and its square
is in arbitrary units of energy, $a.e.$.

\section{Results and Discussion}

\subsection{Speed of drying experiments}
The average of the weight of the bamboo samples as function of drying
time is shown in Fig. \ref{fig:dryingtimes}. Let me focus first on the
black dots. It seems like there is a linear trend in the semi-log
graph for the first three hours of drying. This can be related to a
rate process, which can be the removal of the free water (filling the
interior of the cells, not chemically bounded). This is followed by a
slowest drying phase, between hours three and eight. This may be due
to the removal of the bound water (from inside the cell walls,
chemically bounded). The thicker samples (red diamonds) only show the
first stage, but the moisture content has not fallen below 2\%, so
there is still bond water in the sample.

\subsection{Energy release histograms and waiting times}
The study of fracture phenomena can be tackled from different
disciplines. From the point of view of critical phenomena and
statistical mechanics, fracture is seen as a series of bursts; whose
size and temporal evolution characterizes the phenomena. Recently
Papanikolau et. al. have taken the Barkhausen noise and studied the
functional form of the noise emitted by the avalanches
\cite{papanikolaou2011universality}. Besides the signal average shape,
they present voltage distributions, power spectrums, and distribution
of avalanche sizes $S$ and durations $T$. Their avalanche size
histogram shows a power law distribution at small sizes, followed by a
tail.
\begin{figure}[htb] 
\begin{center}
\resizebox{7cm}{!}{\includegraphics{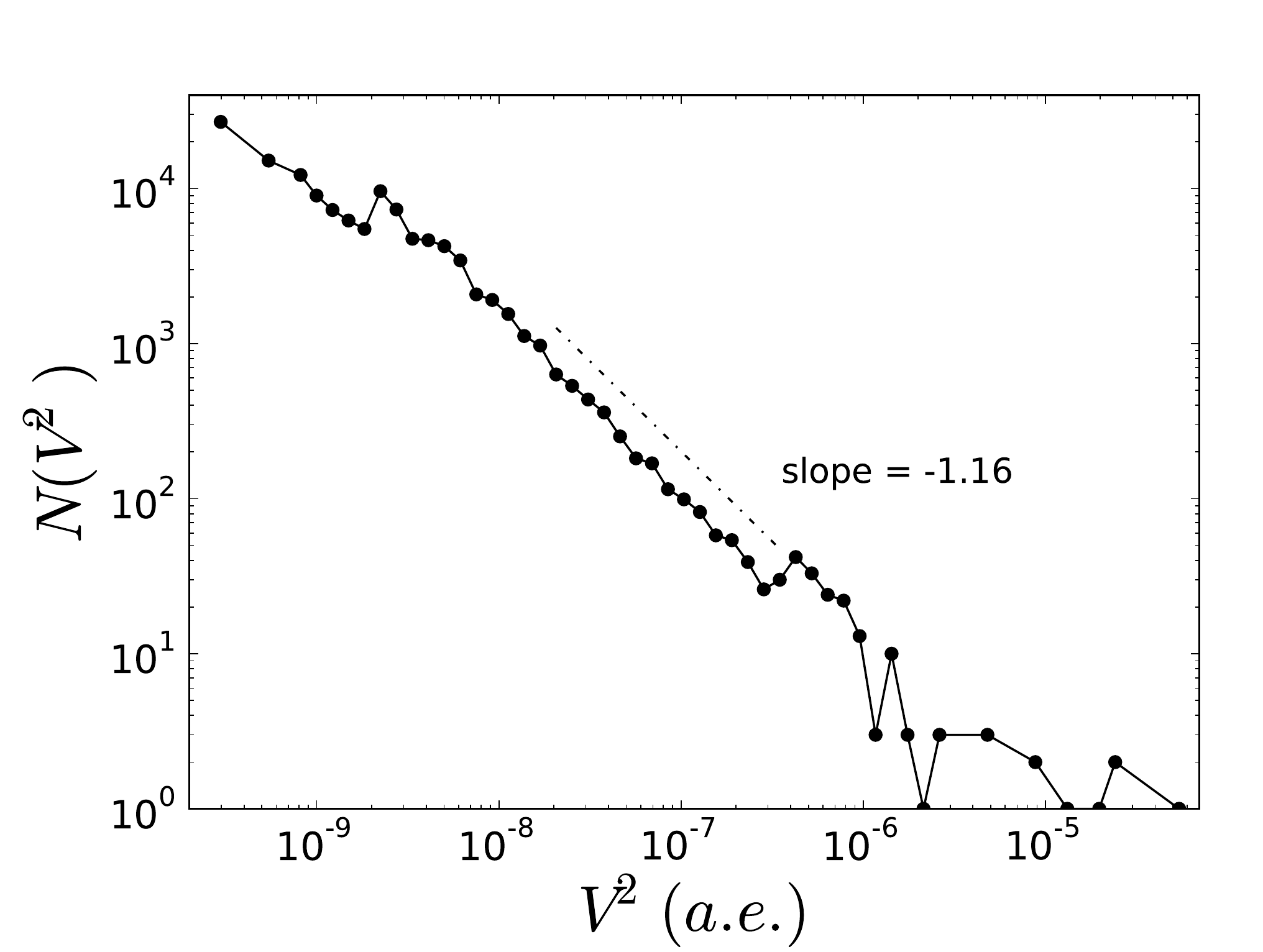}}
\caption{Experimental squared voltage size distributions $N(\Delta)$. The line drawn has a slope of $-1.16$. }
\label{fig:VoltageSizeDistributions}
\end{center}
\end{figure}

In the present case, Fig. \ref{fig:VoltageSizeDistributions} shows the
probability distribution of the square of the voltage recorded by the
microphone, which is proportional to the energy liberated in the
fracture. A power law distribution of avalanche sizes is one
indicative of scale free behavior, as seen for instance in the
Gutenberg-Richter law of avalanches, and in the Barkhausen noise; even
though the mechanisms that generate the fractures at different scales,
there is not a single size of fracture that rules the breaking
phenomena. In this case the distribution of the
square of the sizes follows a power law (which is proportional to the energy released in
the fracture event),
\begin{equation}
  P(\Delta) \propto \Delta^{-\alpha},
\end{equation}
with an exponent of $\alpha = -1.16$. It is interesting to notice that
the power law seems to fit for three decades. In the present system
small fractures and large ones correspond to different kind of
tissues, and can be caused by accumulation or by direct strain; so it
is not at all self evident that the fracture distribution has to be a
power law.

\begin{figure}[htb]
\begin{center}
\resizebox{7cm}{!}{\includegraphics{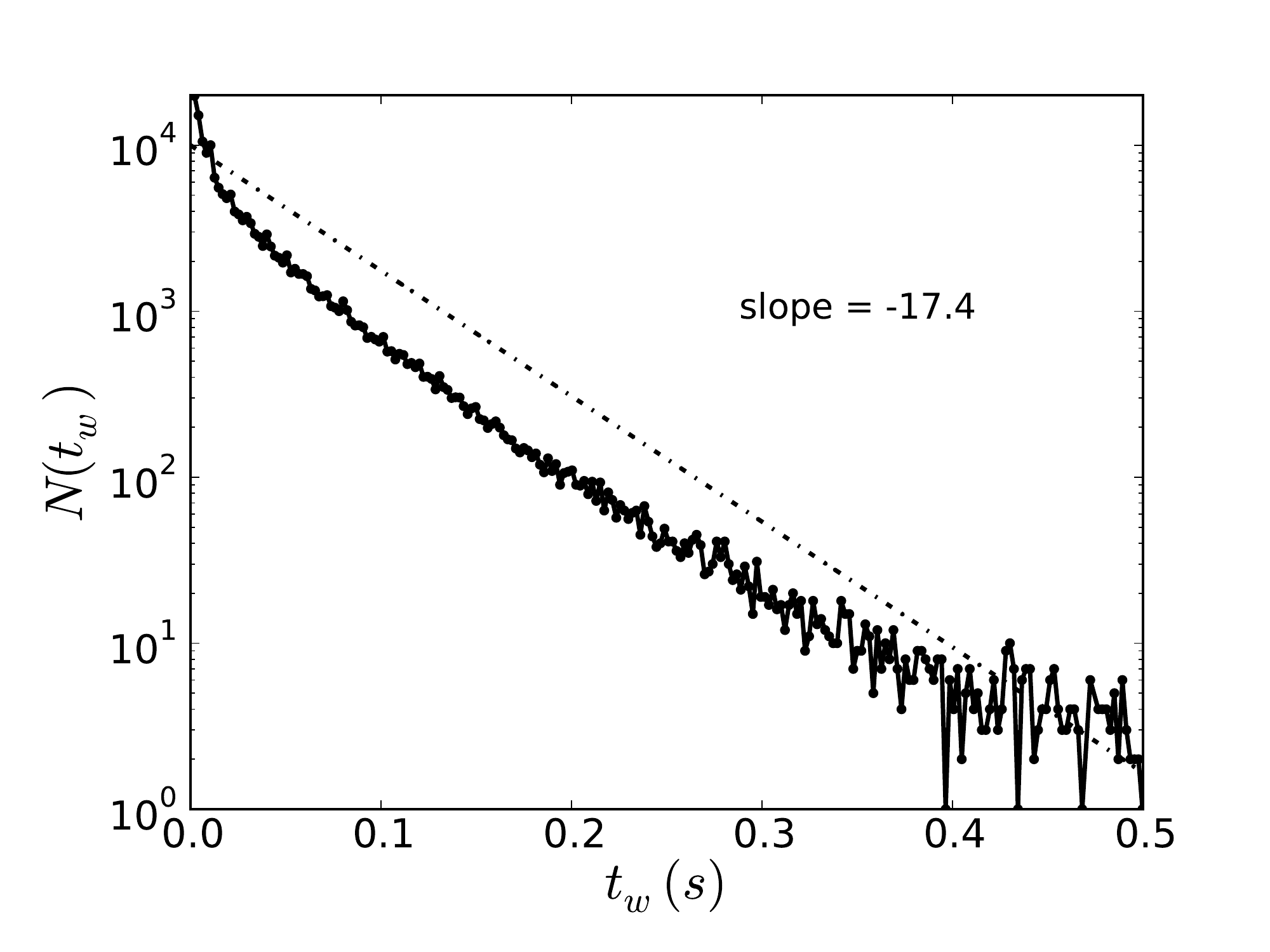}}
\caption{Experimental waiting times between consecutive avalanches.}
\label{fig:ExperimentalWaitingTimeDistribution}
\end{center}
\end{figure}
A second characteristic measurement that can give insight into avalanche processes
is the waiting times between avalanches. For earthquakes this
distribution is known to follow a unified scaling law, mostly a
power-law with a decaying exponential tail
\cite{PhysRevLett.88.178501}. In the present study, on the contrary,
waiting times between avalanches show an exponential decay behavior;
as can be seen in
Fig. \ref{fig:ExperimentalWaitingTimeDistribution}. In this case the
time constant is $0.0574$ $s$. The difference with the earthquake case can
be explained from the fact that in this system the allowed number of
cracks is reduced as there are a finite number of elements to crack or
delaminate; while for earthquakes the possible number of rearrangements
and fractures is orders of magnitude larger. In this way, this system is closer to the granular materials.  In a
recent paper, Michlmayr and Or have studied the relation between the
grain-scale mechanical interactions in sheared granular materials and
the generated acoustic emission characteristics
\cite{michlmayr2014mechanisms}. Their implementation of a damage
accumulation fiber bundle model (similar to the one by Kun et. al),
shows an exponential decay of the number of events as function of the
energy (their Fig. 3, inset); which is the analogous of the waiting
times for that system.

\subsection{structure of energy release}
\begin{figure}[htb]
\begin{center}
\resizebox{7cm}{!}{\includegraphics{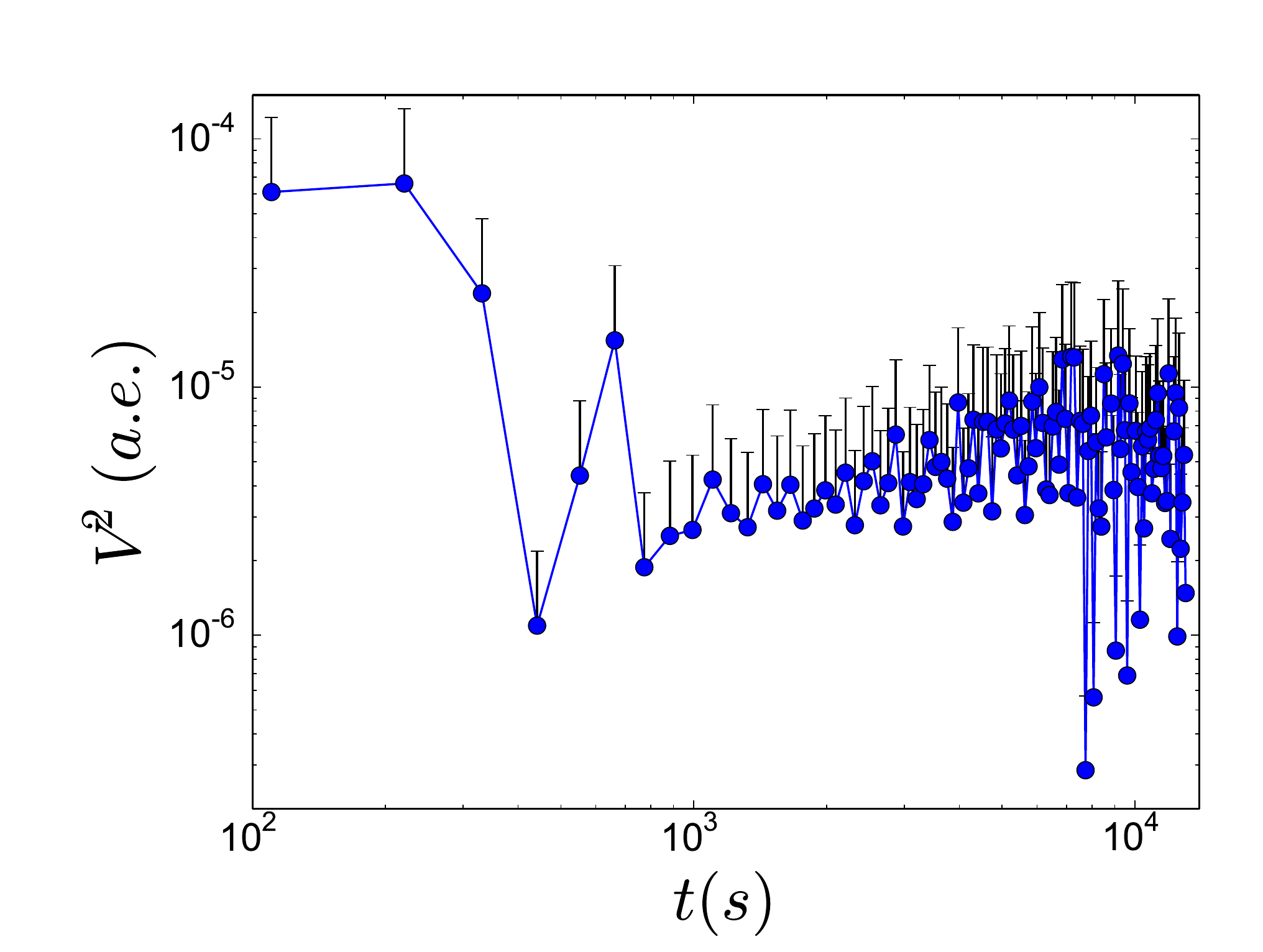}}
\caption{(Color online) Experimental averaged energy release per avalanche as function of time. I only show the upper error bars.}
\label{fig:ExperimentalVoltageAFOClicktime}
\end{center}
\end{figure}

To further characterize the structure of the avalanches, I want to know when do the big avalanches occur with respect to the start of the drying process; as well as whether large energy release indeed means long avalanches. 
For the former question, to identify the evolution of energy release, the size of the avalanche
is averaged over 5 second sound intervals and plotted as function of
the number of the sound interval; and can be seen in
Fig. \ref{fig:ExperimentalVoltageAFOClicktime}. It is possible to identify
two stages, early and late, separated by the sound interval number
$1000$. In the first stage the intensity is a large and decreasing quantity.  At the late stage, there is a steady increase of the size of the avalanches as a function of time. The black lines represent the error bars. 

A possible explanation for the shape of this graph: large values for
short times, followed by a steady increase is this: Initially, the
drying process generates large fractures; creating free surfaces in a
fast process. The presence of this free surfaces hinders the
appearance of large cracks. After this follows a second phase, in
which further fractures may happen mainly for stress
accumulation. This two behaviors are seen in other computational
models \cite{PhysRevE.84.041114}. In that previous work, that I did with other collaborators,
we shown
how the statistics of bursts of breaking events in the case of
shrinkage of a thin layer of 2D material could suggest the presence of
an underlying critical point. In that case, the two states are related
to the way the fracture is created, either by coalescence of
micro-cracks or by the appearance of a large isolated crack that
spanned the sample. 

\begin{figure}[htb]
\begin{center}
\resizebox{7cm}{!}{\includegraphics{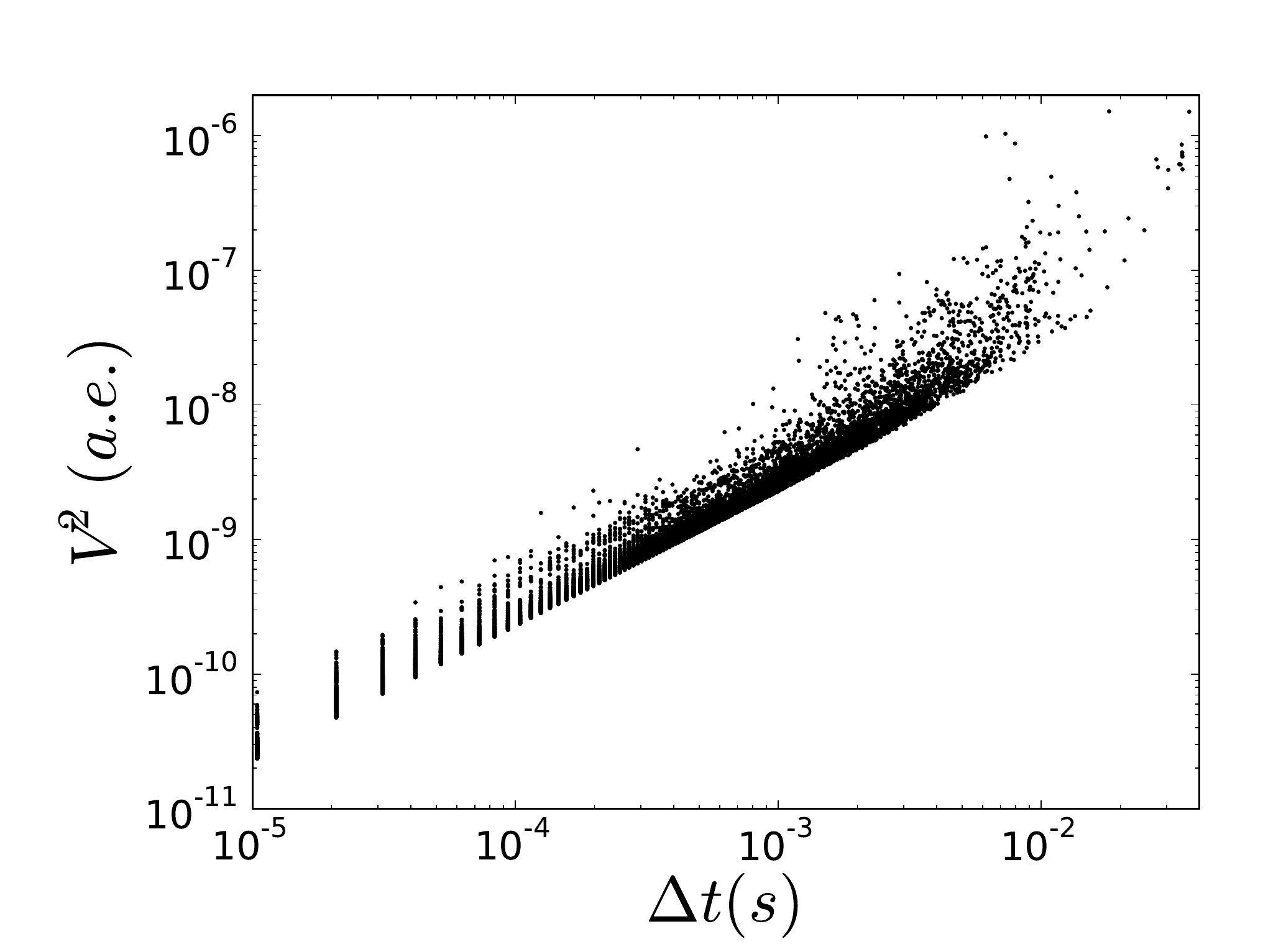}}
\caption{Experimental scatter plot of energy as function of
  avalanche time span.}
\label{fig:ExperimentalEnergySpanScatter}
\end{center}
\end{figure}

To show the relationship between the size of the avalanches and its duration,
I present the correlation between the amount of released energy
in an event and its time span in
Fig. \ref{fig:ExperimentalEnergySpanScatter} as an scatter plot. The
lower cutoff comes from the imposed threshold of measurement.  The
discreetness in time span values is a consequence of the sound
sampling rate of the acquisition transducer. Clearly, there is more
dispersion on the crack events that take longer and release a larger
amount of energy, while the distribution narrows for shorter times. Furthermore it is clear that the relationship between energy release and time span can be fitted to a power law spanning several decades.

\subsection{Shape of the acoustic emission events}
A final way to characterize the process is to study the average shape
of the acoustic emission signal, for different sizes. Knowing which
shape do small and large avalanches have can be useful to identify the
kind of process that generates this fractures. As a matter of fact, in
a previous paper \cite{papanikolaou2011universality} showed a temporal
average avalanche shape, rescaled to unit height and duration, has a
parabolic shape for small avalanches, that flattens towards a unit
pulse shape for longer avalanches; in the case of Barkhausen noise.
\begin{figure}[htbp]\centering 
\resizebox{7cm}{!}{\includegraphics{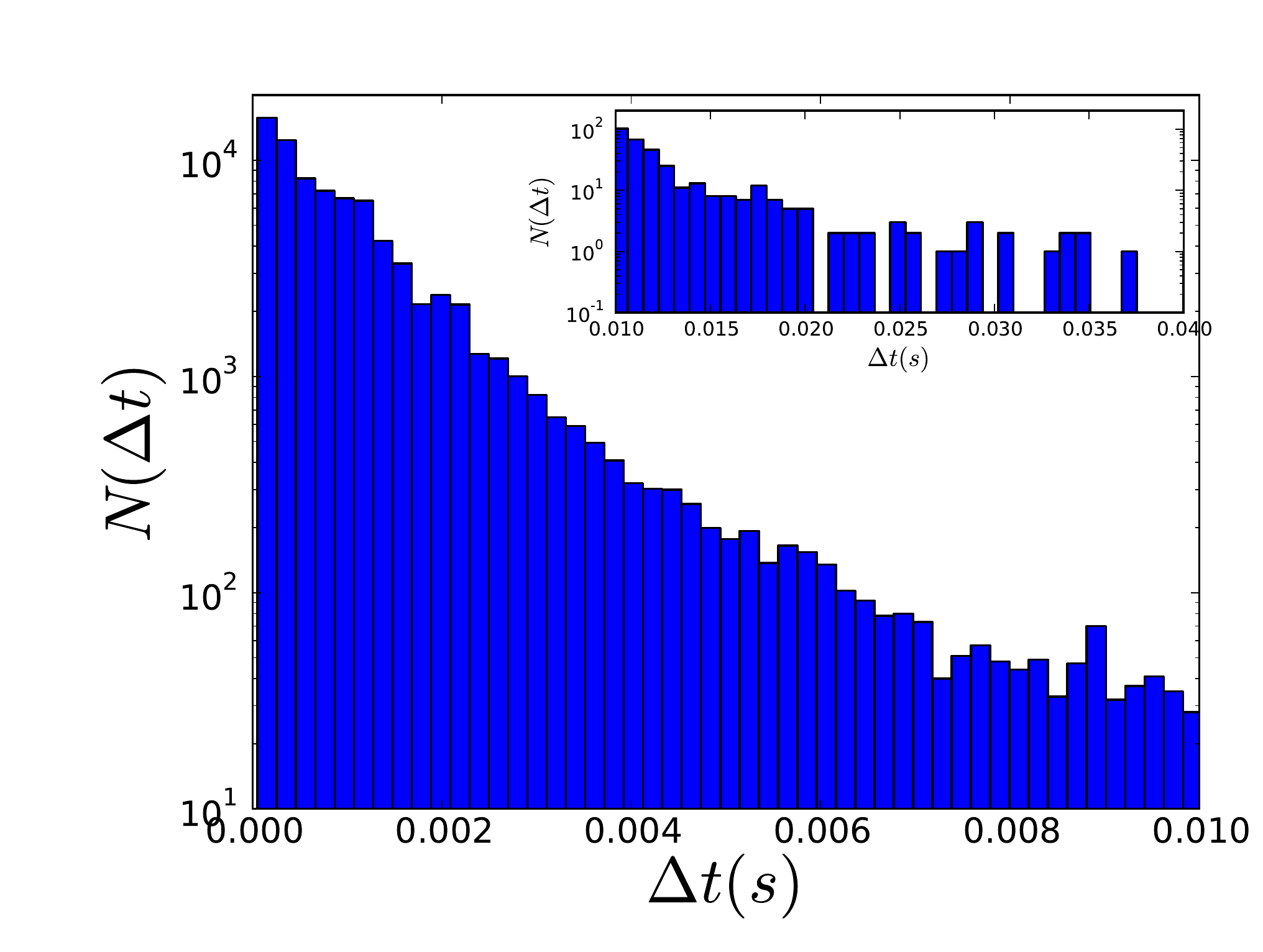}}\\
\caption{\label{HistogramTimes} \small{Histogram of duration of  avalanche events}} \end{figure}

To define the intervals that correspond to avalanches of similar size,
I performed an histogram of the duration of avalanche events,
 Fig. \ref{HistogramTimes}. The number of avalanches varies
over 4 orders of magnitude. In view of this, I hand picked six intervals:
$\Delta t \in (0, 0.002) s$,$\Delta t \in (0.002, 0.005) s$,$\Delta t
\in (0.005,0.01) s$,$\Delta t \in (0.01, 0.012) s$, $\Delta t \in
(0.012,0.02) s$ and $\Delta t \in (0.02, 0.04) s$. Roughly speaking
the first three correspond to the most populous bins ($10^2$ to $10^4$
fractures), short avalanches; and the latter to the large avalanches,
more than $0.01 s$.

\begin{figure}[htbp] \centering 
\resizebox{7cm}{!}{\includegraphics{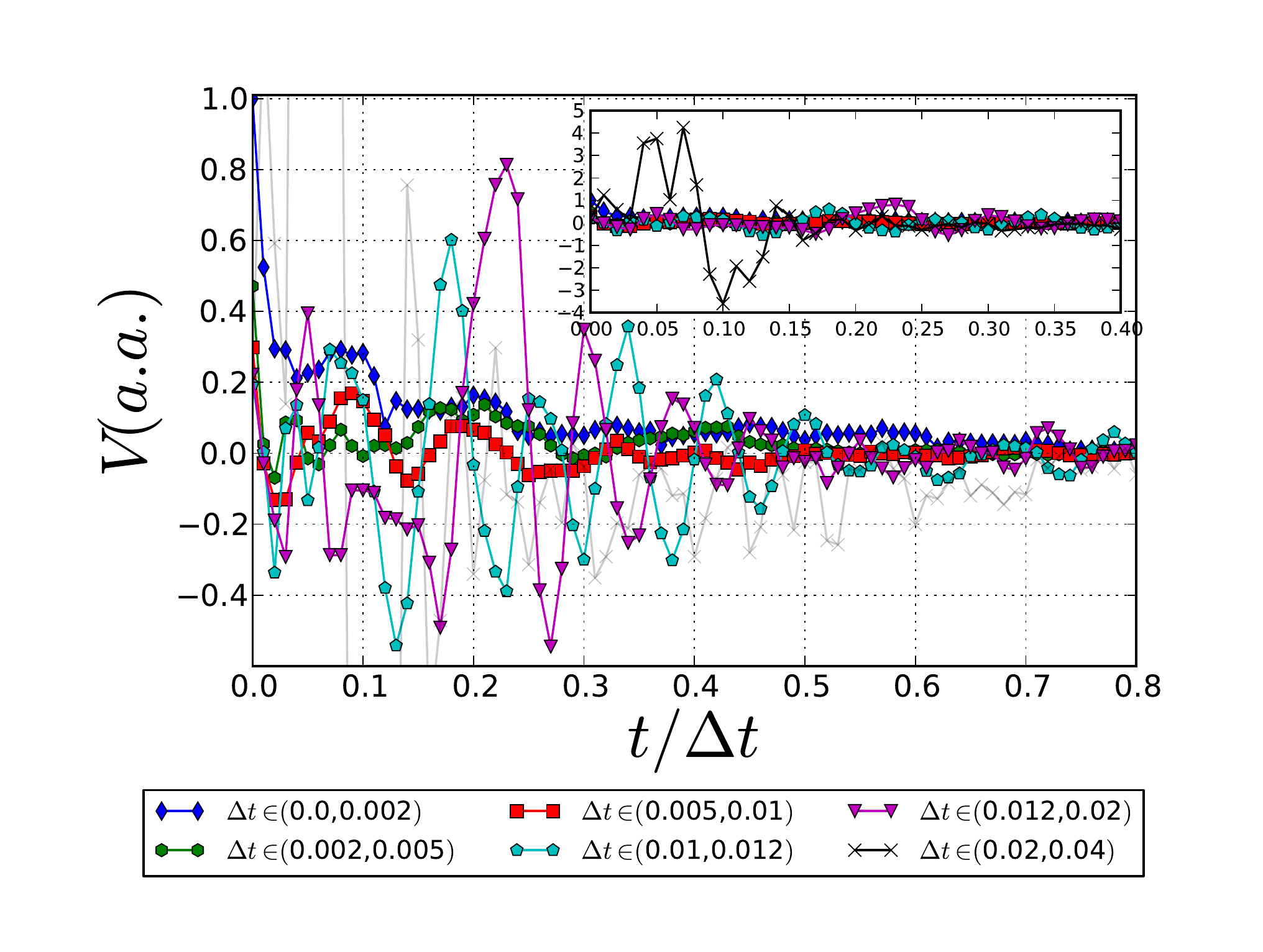}}\\
\caption{\label{acoustic-emission} \small{Normalized averaged signal shape, for 9 different bins of time lengths.}} \end{figure}

\begin{figure}[htbp] \centering 
\resizebox{7cm}{!}{\includegraphics{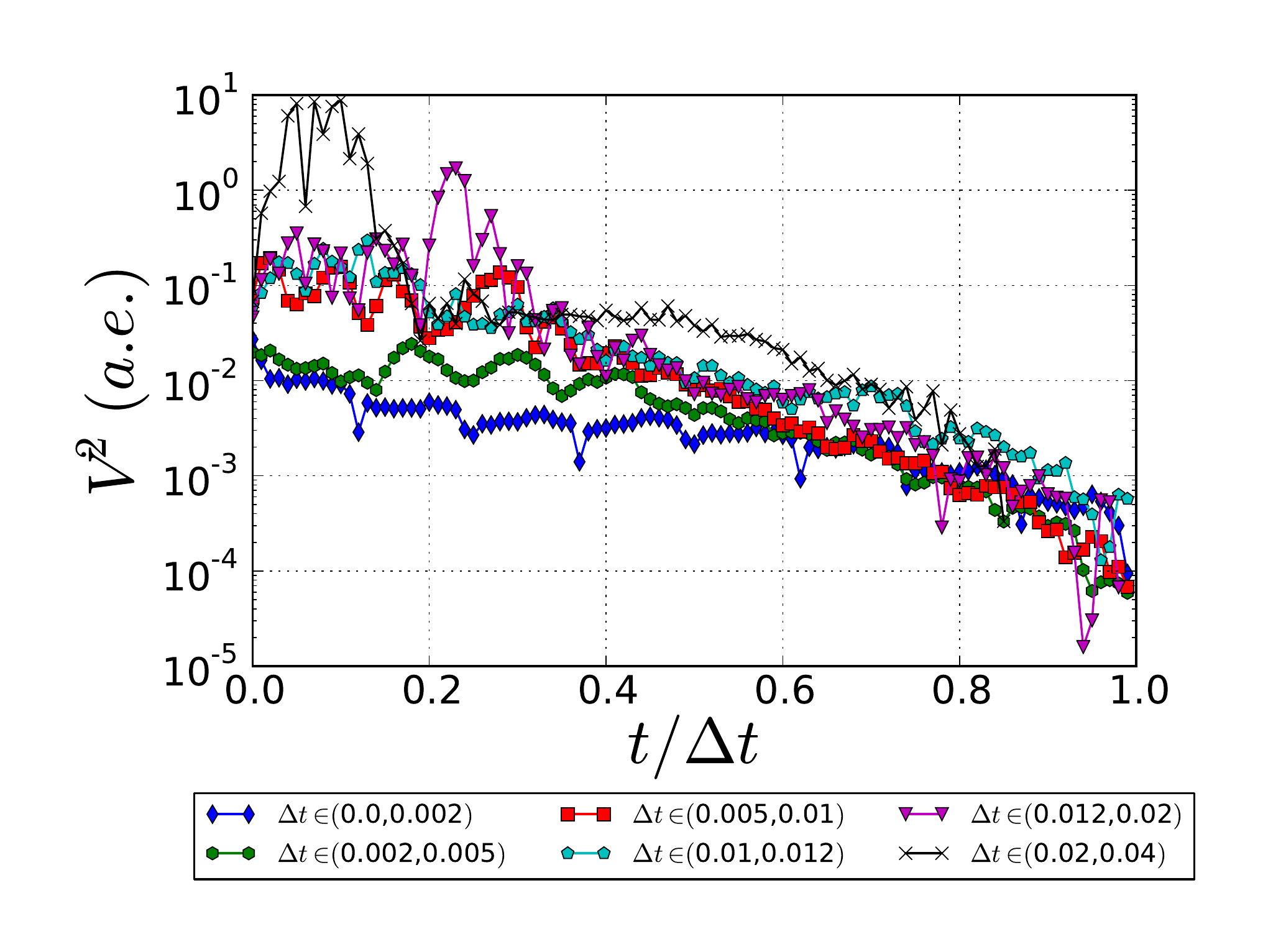}}\\ \caption{\label{acoustic-emision-square-voltage-semilog} \small{Normalized  average square signal shape,for 9 different bins of time lengths.}} \end{figure}

Fig. \ref{acoustic-emission} shows the average avalanche
shape for the different bins, as a function of the normalized
time span. In the main plot I show the first five averaged signals,
with the longest avalanches are shown, but faintly; while in the inset
all the six averaged signals can be seen. Unlike the Barkhausen effect case from
the literature, \cite{papanikolaou2011universality} in which there is
clear symmetry; in this case most of the signals are not symmetrical:
there is a large peak followed by a decay. This can be seen in all
intervals but one ($\Delta t \in (0.012,0.02) s$). In my view, the
asymmetrical shape shape hints for the phenomenon to be that of a
single breaking event, and the decay in the signal corresponds to
attenuation of the noise. Short signals would correspond to small (or
weaker) elements, as parenchymatous tissue; while long signals to
larger (or stronger) elements, as fiber cells. Going back to the
average signal in the range $\Delta t \in (0.012,0.02) s$, the
symmetrical shape here would pinpoint to a different mechanism, namely
the correlated breaking of similarly sized elements; a self induced
growth process in which a breaking triggers similar events. In terms
of the energy, in
Fig. \ref{acoustic-emision-square-voltage-semilog} it can be seen
that the average shape of the largest avalanches has a peak of energy
dissipation which is almost two orders of magnitude larger than that
of the medium and small size avalanches. Would be interesting to test
whether this is caused by the composite nature of the bamboo, or is
rather the result of the dynamics of the avalanche formation; and
would occur also on a more homogeneous material.

\section{Conclusion}
The experimentally measured acoustic emission signals resulting from
drying of bamboo \emph{Phyllostachys Pubescens} shows evidence of
free-scale phenomena. Firstly, the histogram of experimental squared
voltage distributions nicely fit into a power law with exponent of
$-1.16$; while the experimental waiting times do not follow a power
law, rather an exponential of slope $-17.4$ possibly due to the finite
size of the system. Secondly, the scatter plot of energy as function
of avalanche time span also shows a power law that spans over 4 orders
of magnitude. I have also studied the average avalanche shape, and found
that both short and large avalanches have an asymmetrical shape with a
peak at the beginning. This implies that for those cases there is
mainly a single crack event. For medium sized fractures, on the other
hand, I have found a more symmetrical shape; which could be attributed to
fracture growth by a self reinforced process.

The main question that rests to be answered is whether this behavior
is particular to the drying of bamboos; or can be found in other
materials. This can be studied easily from numerical simulations;
which is what I plan to accomplish in the future.

\section*{Acknowledgments}
This work was funded partly by the ``Departamento Administrativo de
Ciencia, Tecnologìa e Innovación de Colombia (COLCIENCIAS)''
(Convocatoria Doctorados Nacionales 2008) and by the Computational
Physics for Building Materials, from Prof. Hans Herrmann at the
Institute for Building Materials, ETH-Z\"urich. I want to thank Ferenc
Kun and Jos\'e D. Mun\~oz for helpful ideas and comments. This work
could not have been done without the help of Miller Mendoza, Julian
Schrenk and Nuno Araujo; with whom I had enlightening conversations
about the system and the measurement process. They also drove me to
pick up the bamboo samples from Frederic Eberts' forest in Chiavari,
Italy (\url{www.bambus.de),} and transported us back to the ETH.


\begin{thebibliography}{10}

\bibitem{PhysRevLett.88.178501}
Per Bak, Kim Christensen, Leon Danon, and Tim Scanlon.
\newblock Unified scaling law for earthquakes.
\newblock {\em Phys. Rev. Lett.}, 88:178501, Apr 2002.

\bibitem{habibi2014crack}
Meisam~K Habibi and Yang Lu.
\newblock Crack propagation in bamboo's hierarchical cellular structure.
\newblock {\em Scientific reports}, 4, 2014.

\bibitem{Janssen1981}
Julius~J.A. Janssen.
\newblock {\em Bamboo in building structures}.
\newblock PhD thesis, Eindhoven University, 1981.

\bibitem{michlmayr2014mechanisms}
Gernot Michlmayr and Dani Or.
\newblock Mechanisms for acoustic emissions generation during granular
  shearing.
\newblock {\em Granular Matter}, 16(5):627--640, 2014.

\bibitem{Montoya06}
Jorge~Augusto Montoya~Arango.
\newblock {\em
  \href{http://ediss.sub.uni-hamburg.de/volltexte/2006/3004}{Trocknungsverfahr%
en f�r die Bambusart Guadua angustifolia unter tropischen Bedingungen}}.
\newblock PhD thesis, Universit�t Hamburg, Von-Melle-Park 3, 20146 Hamburg,
  2006.

\bibitem{Obataya05B}
Eiichi Obataya, Joseph Gril, and Patrick Perr\'e.
\newblock Shrinkage of cane (arundo donax) ii. effect of drying condition on
  the intensity of cell collapse.
\newblock {\em Journal of Wood Science}, 51(2):130 -- 135, 2005.

\bibitem{Obataya05A}
Eiichi Obataya, Joseph Gril, and Bernard Thibaut.
\newblock \href{http://dx.doi.org/10.1007/s10086-003-0578-y}{Shrinkage of cane
  ( \textit{Arundo donax}) I. Irregular shrinkage of green cane due to the
  collapse of parenchyma cells}.
\newblock {\em Journal of Wood Science}, 50:295--300, 2004.
\newblock 10.1007/s10086-003-0578-y.

\bibitem{papanikolaou2011universality}
Stefanos Papanikolaou, Felipe Bohn, Rubem~Luis Sommer, Gianfranco Durin,
  Stefano Zapperi, and James~P Sethna.
\newblock Universality beyond power laws and the average avalanche shape.
\newblock {\em Nature Physics}, 7(4):316--320, 2011.

\bibitem{VILLA12}
Gabriel Villalobos.
\newblock {\em A statistical model of fracture due to drying in Bamboo Guadua
  Angustifolia}.
\newblock PhD thesis, Universidad Nacional de Colombia, 2012.

\bibitem{PhysRevE.84.041114}
Gabriel Villalobos, Ferenc Kun, and Jos\'e~D. Mu\~noz.
\newblock \href{http://link.aps.org/doi/10.1103/PhysRevE.84.041114}{Effect of
  disorder on temporal fluctuations in drying-induced cracking}.
\newblock {\em Phys. Rev. E}, 84:041114, Oct 2011.

\bibitem{Yu08}
H.~Q. Yu, Z.~H. Jiang, C.~Y. Hse, and T.~F. Shupe.
\newblock \href{http://www.srs.fs.usda.gov/pubs/31456}{Selected physical and
  mechanical properties of moso bamboo (\textit{Phyllostachys pubescens})}.
\newblock {\em Journal of Tropical Forest Science}, 20(4):258--263, 2008.

\bibitem{YU07}
Yan Yu, Benhua Fei, Bo~Zhang, and Xiang Yu.
\newblock \href{http://swst.metapress.com/content/p7h2137254040278}{Cell-Wall
  Mechanical Properties of Bamboo Investigated by In-Situ Imaging
  Nanoindentation}.
\newblock {\em Wood and Fiber Science}, 39(4):527 -- 535, October 2007.

\bibitem{Zou20091375}
Linhua Zou, Helena Jin, Wei-Yang Lu, and Xiaodong Li.
\newblock
  \href{http://www.sciencedirect.com/science/article/B6TXG-4TYR05H-1/2/ef0811d%
9d8867a3378f5b2aa9ba8759c}{Nanoscale structural and mechanical characterization
  of the cell wall of bamboo fibers}.
\newblock {\em Materials Science and Engineering: C}, 29(4):1375 -- 1379, 2009.

\end{thebibliography}

\addcontentsline{toc}{chapter}{Bibliography} 
\bibliographystyle{plain}

\end{document}